 \documentclass[prb,twocolumn,showpacs]{revtex4-1}

\usepackage{graphicx}      
\usepackage{dcolumn}       
\usepackage{bm}
\usepackage{amssymb}            

\begin{document}

\title{Single- and double-wall carbon nanotubes fully covered with tetraphenylporphyrins:
Stability and optoelectronic properties from {\it ab-initio} calculations}

\author{Walter Orellana}
\affiliation{Departamento de Ciencias F\'{\i}sicas, Universidad Andres Bello, 
Avenida Rep\'ublica 220, 837-0134, Santiago, Chile}


\begin{abstract} 
The optoelectronic properties of single- and double-wall carbon nanotubes (CNTs) noncovalently 
functionalized with tetraphenylporphyrins (TPPs) are addressed by dispersion-corrected {\it ab initio} 
calculations. Five CNT species with different chiralities were considered. We find that the most stable 
configurations are those where the CNTs are fully covered by TPPs, exhibiting binding energy of about 
2~eV/TPP. The semiconducting CNT-TPP compounds show optical response characterized by a strong 
absorption associated to the TPP bands, with increasing intensity with the TPP concentration. In addition, 
molecular dynamic simulations show that the compounds would be stable at temperatures as high as 
100$^{\circ}$C.
\end{abstract} 

\maketitle

\section{Introduction}
The noncovalent functionalization of carbon nanotubes (CNTs) with porphyrin chromophores has 
proven to be an efficient method that combine the strong photoabsorption of porphyrin with the CNT 
electron transport and charge separation properties. This supramolecular assembly has been explored 
recently as light-harvesting donor/acceptor complexes~\cite{Guldi_05,Hecht_06,Mhuir_06, Casey_08, 
Garrot_11,Roquelet_12,Gupta_11,Sprafke_11,Zhong_13,Zhang_14}, as well as for the electrocatalysis of
the oxygen reduction reaction~\cite{Zagal_12,Orellana_11,Orellana_12,Orellana_13}.
Some schemes proposed for the such a functionalization include the CNT encapsulation in micelles of 
surfactant~\cite{OConnell_02} or in polymers. Recently, CNT-porphyrin complexes 
obtained from a micelle-swelling technique have shown an extremely efficient energy transfer from 
the photosensitized porphyrin to the nanotubes with a remarkable stability, confirming the potentiality 
of these complexes for light-harvesting applications~\cite{Roquelet_10}. However, some key parameters
like the TPP concentration on the CNT surface and the stability of the compounds are not yet clarify.

Another important issue concerning the noncovalent functionalization of CNTs with organic molecules
or polymers is the ability to discriminate between nanotube species in terms of either diameter or chiral 
angles. Experiments have reported a substantial separation of single-wall CNT according to their chirality 
(metallic versus semiconducting) after the adsorption of porphyrins and organic polymers~\cite{Li_04,Chatto_02}. 
Other experiments reported  improvement in the preparation of CNT-polymer solutions with a high degree 
of selectivity, favoring CNTs with large chiral angles and the inhibition of the metallic species.\cite{Nish_07} 
Although the mechanism causing the selectivity is not clearly understood, the above experiments suggest 
that the CNT-polymer bonding is strongly influenced by both the relative orientation of the polymer chain to 
the nanotube structure, as well as possible charge transfer from metallic CNTs.

In the present work we investigate semiconducting single- and double-wall CNTs noncovalently functionalized 
with free-base tetraphenylporphyrin (TPP) molecules. Our goal is to give insight in the stability of such 
complexes with temperature and also determine its electronic properties and optical response. Additionally, 
we consider both metallic and semiconducting CNTs with a similar diameter in order to identify a possible 
chirality dependence in the TPP adsorption strength. Our results show that the TPP molecules tends to 
cover the whole CNT surface, driven by quite strong dispersive forces. The CNT-TPP binding energies would 
depends on the CNT diameters instead of chirality, as previously found for a single TPP adsorbed on different 
CNTs~\cite{Orellana_15}.
We also find that the CNT-TPP complexes would be stable at temperatures as high as 100$^{\circ}$C according 
to our molecular dynamic simulations. Interestingly, the TPP absorption bands remain almost unchanged after 
the TPP stacking on the CNT surface, showing a small redshift. However, the intensities of the TPP absorption 
bands increase with the concentration of molecules on the CNTs, independently if they have single or double wall. 

\section{Computational Method}
The calculations are carried out using density functional theory (DFT) including Van der Waals (VdW) 
dispersion-corrected exchange-correlation functional, as implemented in the SIESTA {\it ab initio} package
\cite{Siesta_02}. A double-$\zeta$, singly polarized basis of localized atomic orbitals is used, while the 
core-electron interaction is described with norm-conserving pseudopotentials. We investigate the TPP 
adsorption on the metallic (10,10) and semiconducting (16,0)  single-wall CNTs, with diameters of 13.9 
and 12.8~\AA, respectively. We also consider the double-wall CNTs (10,10)@(5,5) and (16,0)@(8,8), 
where the inner CNTs have diameters of 7.0 and 6.4~\AA, respectively. These CNTs can host up to 
five TPP molecules around the perimeter. In order to describe the CNTs fully covered  with a single layer of 
TPPs, we use a supercell with a volume of ($38 \times 38 \times a_0$)~\AA$^3$,  and periodic boundary 
conditions along the nanotube axis ($z$ direction). For (16,0) and (10,10) CNTs, $a_0$ is chosen to be 
four and seven times de CNT lattice constant, of 4.334 and 2.500~\AA, respectively. 
For the double-wall CNTs, the $a_0$ parameter is the same. The cell size along $z$ can host one molecule, 
giving a separation between images of about 3.8~\AA. These supercells allow us to describe the 
(10,10)-5TPP and (16,0)-5TPP compounds with a maximum concentration of about 0.3~TPP/\AA. 

We also study the (6,5) CNT fully covered with a single layer of TPPs. This CNT has a diameter of 7.7~\AA\ 
and can host up to four molecules around the perimeter, as previously established~\cite{Orellana_14}. To 
describe the CNT periodicity along $z$, we use a supercell with a volume of ($30 \times 30 \times a_0$)~\AA$^3$, 
where $a_0$ is the CNT lattice constant, of 41.327~\AA. The cell size along $z$ can host a maximum of two 
molecules side by side, therefore with this supercell we can describe two adjacent rings of TPPs around 
the CNT, that is the (6,5)-8TPP compound. This configuration represents the full surface coverage with a 
concentration of about 0.2 TPP/\AA. 

The self-consistency of the density matrix is achieved with a tolerance of 10$^{-4}$  and a common basis 
size or energy shift of 10~meV is applied. A grid cutoff of 100~Ry and the $\Gamma$ point were used for 
the real-space and $k$-space integration, respectively. The accuracy of these parameters were tested 
considering larger grid cutoff of 150~Ry and $k$ mesh of $1\times 1\times 10$. Negligible variation in total 
energy, optical spectra and band structures were found, ensuring converged results. For the optical calculations, 
we use a $1 \times 1 \times 31$ $k$-point mesh, for incident light polarized along the CNT axis. The CNT-TPP 
binding energy is calculated by the energy difference between adsorbed and separated constituents, considering 
corrections due to the basis set superposition error. The CNT-TPP systems were fully relaxed by conjugate 
gradient minimization until the forces on the atoms were less than 0.05~eV/\AA. The dynamical properties and 
stability of the (16,0)-TPP, (10,10)-TPP and (6,5)-TPP systems were investigated by molecular dynamic 
simulations in the $NVT$ ensemble using the Nos\'e-thermostat approach at 373 K over a total simulation time 
of 4~ps, with a time step of 1~fs.

\begin{figure}[ht]
\center
\includegraphics[width=8cm]{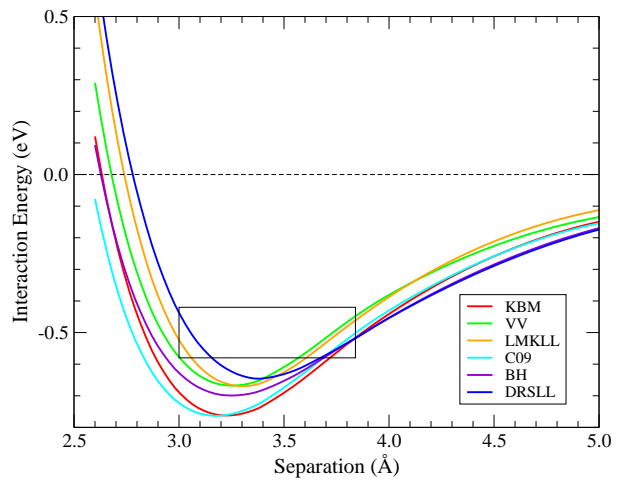} 
\caption{Interaction energy as a function of the separation for the benzene adsorption 
on graphene calculated with different parametrizations for the Van der Waals density 
functional. The square box indicates the available experimental data.}
\label{f1}
\end{figure}

The physisorption of the TPP molecules on the CNT surfaces is assessed by VdW density functional. We test 
different parametrizations for the VdW functional by calculating the benzene adsorption on graphene. These 
parametrizations include those of
Dion {\it et al.} (DRSLL)~\cite{Dion_04}; 
Klimes {\it et al.} (KBM)~\cite{Klimes_09};  
Vydrov and Van Voorhis (VV)~\cite{Vydrov_10};
Cooper (C09)~\cite{Cooper_10};
Berland and Hyldgaard (BH)~\cite{Berland_14}; and
Lee {\it et al.} (LMKLL)~\cite{Lee_10}.
Figure~\ref{f1} shows our results for the interaction energy as a function of the adsorption
distance for the different VdW parametrizations. The square box indicates the experimental 
binding energy of benzene on graphite of $0.5 \pm 0.08$~eV~\cite{Zacharia_04}, whereas 
the variation in distance ranges between the experimental interlayer separation of graphite 
(3.35~\AA) and that of double layer graphitic carbon (3.84~\AA)~\cite{Iijima_79}. As can be 
seen in Fig.~\ref{f1}, the parametrization that best approaches to the experimental binding 
energy of benzene on graphite is that of DRSLL. The difference with respect to the experimental 
results is about 0.15 eV, which in some way represent the error of the binding energy calculations. 
The VdW parametrization of DRSLL will be used throughout this work.

The optical response of the functionalized CNTs is obtained through first-order
time-dependent perturbation theory, by calculating the imaginary part of the dielectric 
function ($\varepsilon_2$). $\varepsilon_2$ gives us a first approach for the optical 
absorption coefficient and it is calculated according to the equation:
\begin{equation}
\varepsilon_2(\omega) = A \int d\mathbf{k}\sum_{c,v} |\hat{\epsilon}\cdot  \langle
\Psi_c({\bf k})|{\bf r}|\Psi_v({\bf k}) \rangle|^2\, \delta (E_c- E_v-\hbar\omega),
\label{e1}
\end{equation}
where $A$ is a constant that depends on the cell size; $\Psi_c$ and $\Psi_v$ are the occupied 
and empty Kohn-Sham orbitals, respectively. The delta function represents the conservation of 
energy, which is described by a gaussian function with a smearing of 0.06~eV. We must note that 
DFT calculations do not include electron-hole and electron-electron interactions, the so called 
many-body effects.  An accurate description of the optical response or photoexcitations in CNT-TPP 
compounds needs theories beyond DFT. However, these calculations are currently prohibitive 
considering the size of the systems under study. Thus, $\varepsilon_2$ as calculated by 
perturbation theory can give a first approach for the CNT-TPP absorption spectra, which are 
typically redshifted by about 0.8~eV with respect to those including many body effects.  

\section{Results and Discussions}
\subsection{Stability and electronic properties}

\begin{figure}[ht]
\center
\includegraphics[width=8.0cm]{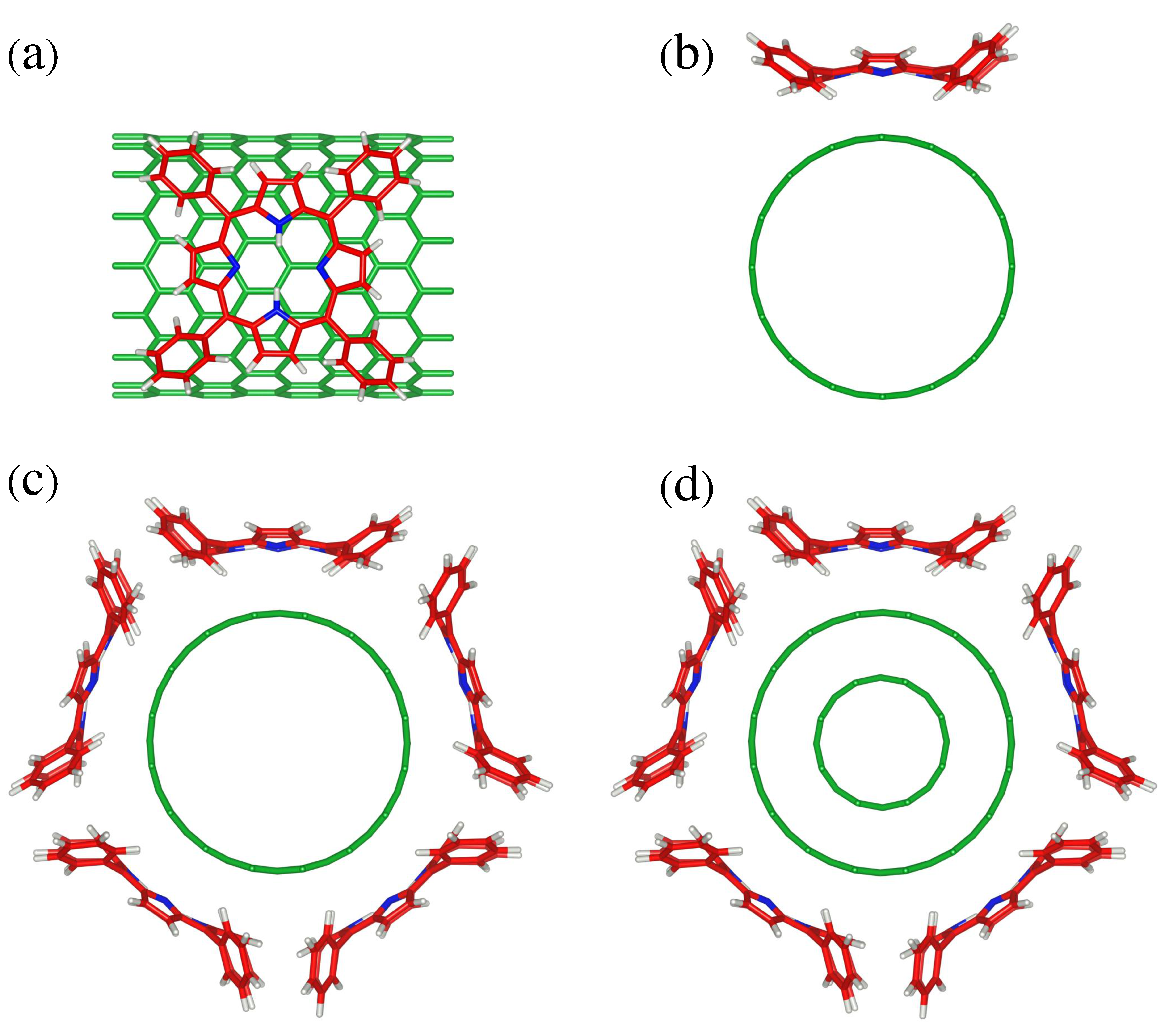} 
\caption{Schematic representation of TPP molecules adsorbed on single- and double-wall CNTs in the 
equilibrium geometry. (a)-(b) Top and front views of (16,0)-1TPP, (c) (16,0)-5TPP, and (d) (16,0)@(8,0)-5TPP.}
\label{f2}
\end{figure}

Figure~\ref{f2} illustrates the TPP molecules adsorbed on the semiconducting (16,0) CNT in the 
equilibrium geometry. The molecules are initially placed in positions close to those depicted in 
Fig.~\ref{f2}, but  with larger separations among them. Then, the whole system is allowed to relax 
without any constraint. The most stable configuration is found for an array of five TPP molecules 
covering the CNT perimeter, forming a ring around the CNT. The system evolves in such a way that 
the TPP molecules tend to fit in the most compact configurations, that is with the phenyl group of one 
molecule between those of the neighboring molecules. The TPP binding energy in the (16,0)-5TPP 
complex is found to be of 2.10~eV/TPP, which is 0.14~eV higher in energy than the adsorption of a 
single TPP, suggesting that the full surface coverage is energetically favorable. 
The TPP adsorption on the double-wall CNT (16,0)@(8,0) shows a binding energy of 2.20~eV/TPP. 
This suggests that the relevant interaction behind the strong adsorption occurs between the adjacent 
aromatic systems with a small contribution from the inner CNT. For the metallic compounds 
(10,10)-5TPP and (10,10)@(5,5)-5TPP, we observe a similar behavior. 

Previously, we reported calculations for the TPP adsorption on the  semiconducting (6,5) CNT, of 7.7~\AA\ 
in diameter, considering different concentrations of adsorbed molecules~\cite{Orellana_14}. 
The most stable compound is found for a configuration of four TPPs covering the CNT perimeter, with a 
binding energy of 2.45~eV/TPP. In the present work we have realized that the TPP binding energy strongly 
depends on the parametrization of the VdW interaction, as shown in the Fig.~\ref{f1}. Thus, we have 
reevaluated the TPP binding energy for the (6,5)-4TPP compound, considering the best  parametrization 
for the VdW density functional. Our results show a lower binding energy of 1.79~eV/TPP, which we believe 
is more realistic than our previous result.

Table~\ref{t1} lists the adsorption distances and binding energies for the TPP molecules adsorbed 
on the single- and double-wall CNTs. 
\begin{table}[ht]
\caption{Adsorption distance ($r$) and binding energy ($E_b$) of TPP molecules
adsorbed on single- and double-wall CNTs. $D$ ($d$) is the diameter of the outer 
(inner) double-wall CNTs.}
\begin{ruledtabular}
\begin{tabular}{lcccc}
Compound & $D$ (\AA) & $d$ (\AA) & $r$ (\AA) & $E_b$ (eV/TPP)\\
\hline
 (6,5)-1TPP          &  7.68   &  $-$   &  3.64  & 1.31  \\
 (6,5)-4TPP          &  7.65   &  $-$   &  3.14  & 1.79  \\
 (6,5)-8TPP          &  7.65   &  $-$   &  3.14  & 1.95  \\
 (16,0)-1TPP         & 12.81   &  $-$   &  3.16  & 1.96  \\
 (16,0)-5TPP         & 12.82   &  $-$   &  3.23  & 2.10  \\
 (16,0)@(8,0)-1TPP   & 12.87   &  6.44  &  3.12  & 1.97  \\
 (16,0)@(8,0)-5TPP   & 12.92   &  6.44  &  3.18  & 2.20  \\
 (10,10)-1TPP        & 13.89   &  $-$   &  3.24  & 2.02  \\
 (10,10)-5TPP        & 13.88   &  $-$   &  3.23  & 2.26  \\
 (10,10)@(5,5)-1TPP  & 13.90   &  7.02  &  3.20  & 2.04  \\  
 (10,10)@(5,5)-5TPP  & 13.92   &  7.02  &  3.21  & 2.24  \\
\end{tabular}
\end{ruledtabular}
\label{t1}
\end{table}
According to this results we can infer some general properties for the CNT-TPP compounds: 
(i) The TPP binding energy tends to increase with the CNT diameter, somewhat expected due 
to the CNT-TPP $\pi$-stacking interaction, which is maximum for planar systems.
(ii) For single- and double-wall CNTs, the TPP binding energy is almost the same, suggesting 
that the adsorption interaction is localized, that is, depending mostly on the adjacent 
aromatic systems.
(iii) The number of molecules on the CNT surface tends to increase the adsorption energy, 
favoring the full surface coverage. We observe an increase in the compound binding energy of 
about 0.2~eV per molecule for higher concentrations. We attribute this result to an additional 
VdW attractive force among the TPP phenyl groups which would play a key role in the stabilization 
of the TPP layer covering the CNT.

\begin{figure}[ht]
\center
\includegraphics[width=7cm]{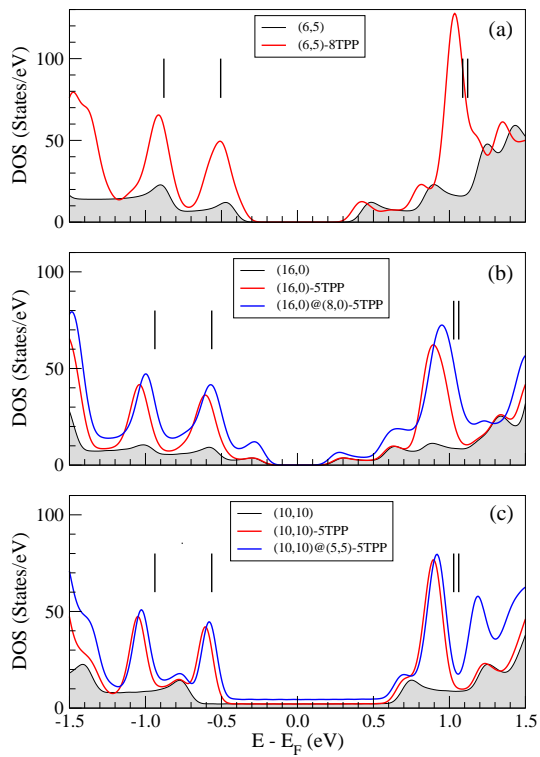} 
\caption{Density of states for the CNT-TPP compounds:
(a) (6,5)-8TPP,
(b) (16,0)-5TPP and (16,0)@(8,0)-5TPP, and
(c) (10,10)-5TPP and (10,10)@(5,5)-5TPP.
The vertical lines represent the frontier orbitals of the free TPP molecule.} 
\label{f3}
\end{figure}

Figure~\ref{f3} shows the density of states (DOS) close to the Fermi level for the single- and 
double-wall CNTs fully covered with TPP molecules. The Fermi energy ($E_F$) is placed at the center 
of the energy gap for the semiconducting systems. The vertical lines in the figure indicate the frontier 
orbitals of the isolated molecule: HOMO-1, HOMO, LUMO, and LUMO+1. The states added to the
CNT DOS by the HOMO-1 and HOMO levels are identify by peaks at around -1.0 and -0.5~eV,
respectively. Whereas the states associated to LUMO and LUMO+1, which are very close in energy, 
appear as a larger single peak at around 1~eV.
The height of these peaks are related to the number of TPP molecules adsorbed on the CNTs.
On the (6,5) CNT we put eight TPPs per supercell, referred as (6,5)-8TPP in the Fig.~\ref{f3}(a). 
Therefore, the molecular peaks are larger than those found in the (16,0)-5TPP and  (10,10)-5TPP 
compounds, containing five molecules, as shown Fig.~\ref{f3}(b) and 3(c), respectively. 
We verify that the strong CNT-TPP interaction induces distortions in the adsorbed molecules, 
resulting in small changes in the position of the energy levels, which explain to some extent the 
width of the absorption peaks and also the decrease in the HOMO-LUMO energy with respect
to the free molecule, as shown in the Fig.~\ref{f3}.

For the (6,5) CNT we observe a reduction in its band gap as induced by the TPP adsorption, as shown 
Fig.~\ref{f3}(a). This behavior can be explained for the position of the TPP HOMO levels which  
coincide with the CNT valence band maximum, inducing the rising of these levels inside the CNT band 
gap, as previously discussed for the (6,5)-4TPP compound~\cite{Orellana_14}. Regarding the double-wall 
CNTs fully covered with TPP molecules, we observe essentially the same electronic properties than 
those of the single-wall CNT, unless a small shift to lower energies which is more pronounced for the 
semiconducting CNT species, as shown  Figs.~3 (b) and 3(c). We attribute this shift to a weaker 
interaction between the molecules and the outer CNT due to a screening effect, which are also expressed 
in the larger TPP adsorption distance, as shown in Table~\ref{t1}.

\subsection{Optical absorption properties}

The optical absorption spectroscopy of the TPP molecule shows resonances at 2.831 and 2.357~eV, 
which are associated to the $B$ (Soret) and the first $Q$ bands, respectively~\cite{Roquelet_10}. 
$B$ and $Q$ bands split in two electronic components, $x$ and $y$, owing to the symmetry breaking 
in the TPP molecule due to the alignment of the N-H bonds in one direction. Theoretically, the optical 
response of the CNT-TPP compounds can be estimated through the imaginary part of the dielectric 
function ($\varepsilon_2$), which is proportional to the optical absorption, considering a light polarization 
parallel to the tube axis.
For the free TPP, our results show clear resonances at 2.04 and 1.60~eV, which we associate to the 
$B$ and $Q$ bands, respectively. These resonances can be characterized by transitions between TPP
frontier orbitals.\cite{Gouterman_63} Consequently, $Q_x$ ($Q_y$) can be associated to the transition 
from HOMO to LUMO (LUMO+1), while $B_x$ ($B_y$) to the transitions from HOMO-1 to LUMO 
(LUMO+1). The close proximity in energy between LUMO and LUMO+1 levels, of less than 0.1~eV, 
leads to the appearance of single peaks in the optical spectrum of the compound, which we called 
$Q_{xy}$ and $B_{xy}$.  We observe that the theoretical energy difference between $Q_{xy}$ 
and $B_{xy}$ bands (0.44~eV) is close to those observed in experiments (0.47~eV)~\cite{Roquelet_10}. 
Therefore we can say that the difference in energy between $Q_{xy}$ and $B_{xy}$ is well represented 
by the frontier orbital approximation~\cite{Gouterman_63}. 
However, experiments show the B band at 2.831 eV, while in our theoretical approach it is 
found at 2.04 eV. Thus, our single-particle approximation shows a general redshift of 0.8 eV in the 
absorption spectra of the CNT-TPP compound, but qualitatively preserving the characteristic of the TPP
absorption bands. 

It is important to note that excitonic effects alter dramatically the optical spectra in semiconducting CNTs. 
For each peak obtained with the single-particle approximation, it is found a series of visible exciton lines 
when hole-electron interaction is included~\cite{Spataru_04}. Optical excitation of CNT in photoluminescence 
or Raman experiments have confirmed these excitonic states, which for CNTs with diameters around 8 \AA\ 
lie several hundred of meV below the conduction band edge~\cite{Wang_05}. If both electron-electron 
and electron-hole interactions were included in our calculations, no redshift on the CNT optical transition 
should be expected. In addition, because the molecular absorption bands are located far from the CNT 
conduction-band edge, the optical transitions of the adsorbed molecules should not be altered. This has 
been confirmed by optical absorption spectroscopy of isolated TPPs and the CNT-TPP 
compounds~\cite{Roquelet_13}.

\begin{figure}[ht]
\center
\includegraphics[width=7cm]{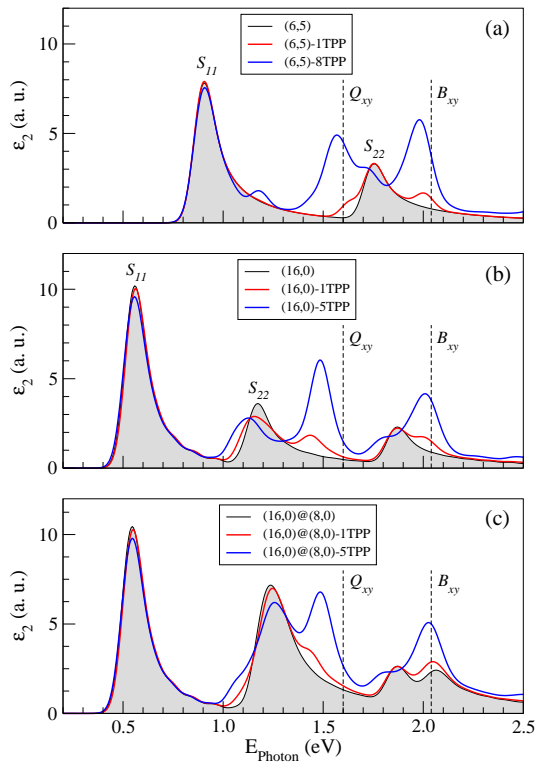} 
\caption{Imaginary part of the dielectric function for the semiconducting CNT-TPP compounds:
(a) The (6,5)-TPP, (b) The (16,0)-TPP, and (c) The (16,0)@(8,0)-TPP. The dashed lines 
indicate the $Q_{xy}$ and $B_{xy}$ absorption bands of the free TPP molecule.}
\label{f4}
\end{figure}

Figure~\ref{f4} shows the imaginary part of the dielectric function for the semiconducting CNTs 
functionalized with TPP molecules with different concentrations, as compared with the pristine CNTs. 
For the fully coverage configurations: (6,5)-8TPP, (16,0)-5TPP and (16,0)@(8,0)-5TPP, we clearly 
identified two peaks at around 1.5 and 2.0~eV which are associated to the $Q_{xy}$ and $B_{xy}$ 
bands of the adsorbed molecules, respectively. The intensity of these peaks can be related 
with the concentration of adsorbed molecules. 
For the (6,5)-8TPP compound, we observe a redshift of the $B_{xy}$   band by about 0.1~eV.
This redshift can be associated to the screening of the Coulomb interactions in the nanotube by the 
adsorbed molecules, in agreement with experimental results~\cite{Roquelet_13}. 
However, the opposite situation occurs for the (16,0)-5TPP and (16,0)@(8,0)-5TPP compounds where
the $Q_{xy}$ band is redshifted while the Soret band remains almost unchanged. Additionally, we note 
that the TPP adsorption on the different CNT-TPP compounds appear to be not affected by the nature
of the CNT, suggesting that multi-wall CNTs would exhibit the same optical properties that those 
observed in single-wall CNTs.

\subsection{Molecular dynamic simulations}
\begin{figure}[ht]
\center
\includegraphics[width=8cm]{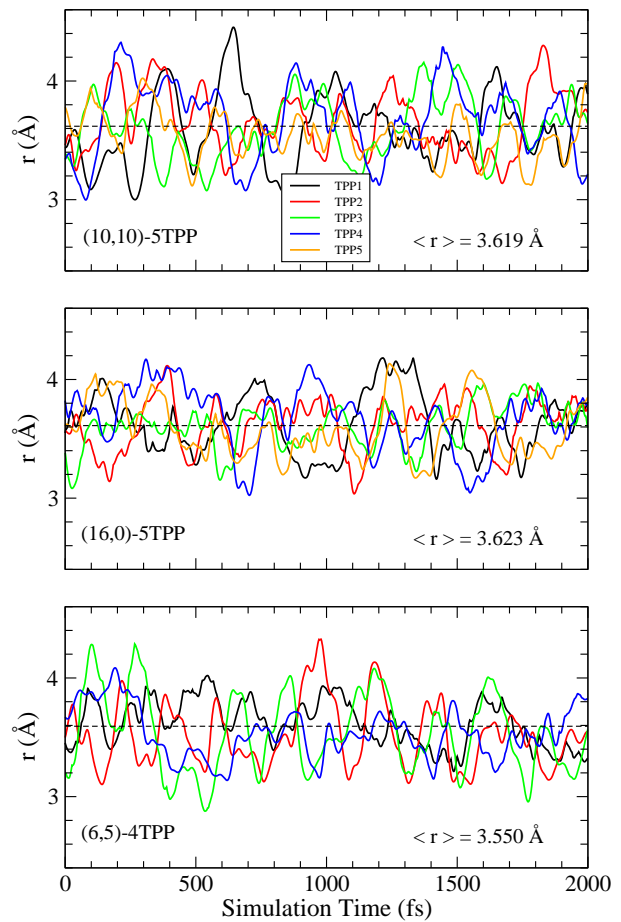} 
\caption{Adsorption distance $r$ as a function of time for the molecular dynamic simulation of the 
CNT-TPP compounds at 373 K. The dashed lines indicate the average adsorption distance. The 
different curves indicate the evolution of $r$ for each TPP molecules.}
\label{f5}
\end{figure}

The stability of the CNT-TPP compounds is investigated through {\it ab initio} molecular dynamic 
simulations at 100$^{\circ}$C, which is a typical temperature for a device operation. We consider 
a total simulations time of 4 ps. After a stabilization during 2 ps, we analyze the adsorption distance 
$r$ of each TPP molecule on the CNT surface for the compounds (10,10)-5TPP, (16,0)-5TPP, and 
(6,5)-4TPP. The evolution of the adsorption distance with time is obtained by taking the minimum 
distance from a TPP N-atom to the closer nanotube C-atom, with the help of the LPMD code 
toolkit~\cite{Davis_10}. 
Our results are shown in Fig.~\ref{f5}. As can be seen, the TPP molecules fluctuate onto the CNT 
surface without desorbing. The mean adsorption distances of the five TPP molecules on the (10,10) 
and (16,0) CNTs are found to be of 3.619 and 3.623~\AA, respectively, which are 11.6\% and 12.2\% 
larger than those found for the equilibrium geometry at 0~K.
The high TPP binding strength is also observed in the smaller (6,5) CNT, with a diameter of 7.7~\AA, 
as shown our MD simulation for (6,5)-4TPP. Here the mean adsorption distance, of 
3.550~\AA, is 13.1\% larger than the adsorption distance at 0~K. This confirms the strong 
attractive interaction of the (6,5)-TPP compounds as experimentally reported~\cite{Vialla_13}. The 
above results suggest that the TPP molecules remain adhered to the CNT surface at temperature
as high as 100$^{\circ}$C. However, the large TPP binding energy, of about 2~eV, suggests 
a much higher desorption temperature. To check this, we performed additional MD simulations for
(16,0)-5TPP, considering a temperature of 500$^{\circ}$C. Our results reveal that the TPP 
molecules still remain adhered on the CNT surface, confirming the high stability of the CNT-TPP
compounds. 

We also note that no sizable changes in the adsorption distance between the metallic and semiconducting 
CNTs is obtained. Considering that the metallic (10,10) and semiconducting (16,0) CNTs have similar 
diameters and they can host up to five TPP molecules around its perimeter with almost the same binding 
energies  (see Table~\ref{t1}), our results suggests that the CNT chirality would not be a relevant 
parameter for the TPP binding strength.

\section{Summary and conclusions}
In summary, using dispersion-corrected DFT calculations and linear optical response we have 
investigated the stability, electronic and optical properties of single- and double-wall CNTs noncovalently 
functionalized with TPP molecules. CNTs with different chiralities were also considered. Our results show 
that the most stable configuration for all CNTs under study are those where they are fully covered by 
TPP molecules. This configuration shows binding energy of around 2~eV/TPP for all the CNT species. We 
find that the molecular aggregation on the CNT surface is an exothermic process with a gain in energy of 
about 0.2~eV. We also find that the complex stability would be independent  of both the CNT chirality 
and the CNT number of walls. Therefore, no evidence of a preferential interaction of TPP molecules with metallic or 
semiconducting CNTs is found. However the CNT diameter appears to be the relevant parameter for the  
increasing TPP binding strength. 

The semiconducting CNT-TPP compounds exhibit optical response characterized by a strong adsorption 
associated to the TPP bands, with increasing intensity with the number of adsorbed molecules. This 
characteristic is preserved in double-wall CNTs, suggesting that the TPP functionalization of multi-wall 
CNTs would exhibit the same optical properties that those observed for the single-wall CNTs.

Finally, the stability of the CNT-TPP compounds is investigated by molecular dynamic simulations. Our 
results show that TPP molecules in the fully covered configurations remain adhered on the CNT surface 
at 100$^{\circ}$C, with adsorption distance fluctuating around 3.6~\AA. 
Further simulations suggest that the CNT-TPP compounds would be stable at temperatures as high as 
500$^{\circ}$C. The above results suggest that semiconducting CNTs fully covered with TPPs would exhibit 
excellent properties as a light-harvesting material. In addition, the stability of the porphyrins/CNT complexes 
is also very important in the field of electrocatalysis as they have shown catalytic activity for the O$_2$ 
reduction reaction~\cite{Orellana_11,Orellana_12}, while hybrid metal-porphyrins/CNT systems are 
currently tested by several groups around the world as electrode materials in fuel cells~\cite{Zagal_12}. 

\acknowledgments
This work  was supported by the funding agency CONICYT-PIA under the Grant Anillo ACT-1107.

\end{document}